\documentclass[pre,twocolumn,showpacs,showkeys,aps]{revtex4}

\usepackage[final]{graphicx}
\usepackage{t1enc}

\begin{document}

\title{\bf The distance between Inherent Structures and the inluence of saddles on
approaching the mode coupling transition in a simple glass former}

\author{Gabriel Fabricius}
\email{fabriciu@fisica.unlp.edu.ar}
\affiliation{Departamento de Física, Universidad Nacional de La Plata, cc 67, 1900
La Plata, Argentina.}
\author{Daniel A. Stariolo\footnote{Research Associate of the Abdus Salam
International Center for Theoretical Physics, Strada Costiera 11, Trieste,
Italy}}
\email{stariolo@if.ufrgs.br}
\homepage{http://www.if.ufrgs.br/~stariolo}
\affiliation{Departamento de Física, Universidade Federal do Rio Grande do Sul,
CP 15051, 91501-970 Porto Alegre, Brazil}

\date{\today}

\begin{abstract} 
We analyze through molecular dynamics simulations of a Lennard-Jones (LJ) binary mixture the
statistics of the distances between inherent structures sampled at temperatures above 
the mode coupling transition temperature $T_{MCT}$. After equilibrating
at $T > T_{MCT}$ we take equilibrated configurations and randomly perturb the
coordinates of a given number of particles. 
After that we find the nearby inherent
structures ($IS$) of both the original and perturbed configurations and evaluate the
distance between them. This distance presents an inflection point at $T_{li} \simeq 1$
with a strong decrease below this temperature which 
goes to a small but nonzero value on approaching $T_{MCT}$.
 In the low temperature region we
study the statistics of events which give zero distance, i.e. dominated by minima, and
find evidence that the number of saddles decreses exponentially near $T_{MCT}$. This
implies that saddles continue to exist even at $T \leq T_{MCT}$. As at  $T_{MCT}$ the
diffusivity goes to zero, our results imply that there are saddles associated with
nondiffusive events at $T < T_{MCT}$.
\end{abstract}

\pacs{61.43.Fs, 61.20.Ja, 61.43.-j}
\keywords{glasses, molecular dynamics, energy landscape}

\maketitle

What is the influence of the potential energy landscape on the dynamical properties
of glass formers? Understanding of this relation is crucial for the development of a
comprehensible thermodynamics of supercooled liquids and glasses. In recent years much
progress have been done in this direction~\cite{debenedetti}. Simulating simple models
of glass formers Sastry et al.~\cite{sastry} found evidence of four regions where the
influence of the landscape is qualitatively different: (a) the high temperature region
where the system is essentially a simple liquid, with exponentially decreasing time
correlation functions and free diffusion,
 (b) a {\em landscape influenced} region characterized
by the onset of nonexponential relaxation for $T_{MCT} < T < T_{li}$, (c) a region
which Sastry et al. called
{\em landscape dominated} in which activated events should be dominant and (d) the
region for $T < T_g$ corresponding to the glass phase. In this work we will be concerned 
with the first two regions and the crossover between the second and the third one.
In regions (a) and (d) the landscape
{\em seen} by the system is essentially flat, of course for very different reasons: in
(a) the energy of instantaneous configurations is much higher than typical potential
energies and in (d) the system is able only to explore a strictly local region mainly
through vibrations and diffusion is avoided. In (b) and (c) the ruggedness of the
landscape has a strong influence on the dynamics and makes a thermodynamic description
very difficult to rationalyze. More recently a more quantitative description of the
role of the landscape has begun to emerge
~\cite{sciortino,angelani,cavagna,middleton,grigera,doye}. 
An interesting way of looking at changes in the landscape topology in the different
regions is to calculate the statistics of saddle points. By looking at the order of
the saddles, i.e. the number of unstable directions $n_s$,
 Angelani et al.~\cite{angelani}
found that this is a well defined function of 
temperature and found evidence that it goes to
zero at a temperature which seems to coincide with the mode coupling transition
temperature $T_{MCT}$~\cite{go.sj.92}.
 The immediate consequence of this result is that the mode 
coupling transition can be interpreted in terms of a topological change in the
landscape: $T_{MCT}$ should mark the transition between a region dominated by saddles
to a region dominated by minima. That this is qualitatively true is now reasonably well
established but the quantitative identification of $T_{MCT}$ with the point where
$n_s$ goes to zero needs to be independently verified for different systems and if 
possible
by looking at several different quantities. 
One of these is the mean squared distance ($MSD$). 
In reference\ \onlinecite{sastry} 
the MSD was calculated between an instantaneous configuration and the
nearby IS. The main conclusion was that at very low temperatures, tipically below 
$T_{MCT}$, it goes linearly to zero as $T\rightarrow 0$, 
an evidence that the $MSD$ comes essentially from harmonic vibrations. 
In reference\ \onlinecite{angelani} it was shown the MSD between saddles and the
corresponding minima suggesting a linear relation also between this MSD and the order
of the saddles $n_s$. 

Here we look at the distances between near inherent structures which give additional
information on the relation between landscape topology and the dynamics of the system.
We performed molecular dynamics simulations on a well known Lennard-Jones binary
mixture of 80 \% particles of type $A$ and  20 \%  particles of type $B$ with 
$\epsilon_{AA} = 1.0$,
$\epsilon_{BB} = 0.5$,
$\epsilon_{AB} = 1.5$,
$\sigma_{AA}   = 1.0$,
$\sigma_{BB}   = 0.88$,
$\sigma_{AB}   = 0.8$ 
at a density of 1.204 that has been extensively studied for 1000 particles
by Kob {\it et al.} \cite{kob1,kob2} who obtained $T_{MCT} \simeq 0.435$.
In the present work we used $N=N_A+N_B=250$.
We first equilibrate the system at temperatures 
 $T=5, 4, 3, 2, 1, 0.8, 0.6, 0.55, 0.5$ and $0.47$. 
We study 32 samples at each temperature following different thermal
treatments in order to have the same statistics than in 
 reference\ \onlinecite{kob1} and verify that extrapolating our data to lower
temperatures the diffusivity should go to zero at the
same value of $T_{MCT}$. 
The distance between IS is calculated as follows: once the system 
is equilibrated at a 
temperature $T$  a typical configuration is taken.
Then we make
a {\em damage} on the equilibrated configuration by changing randomly the coordinates 
of $n_d$ particles chosen also randomly and
the IS corresponding to the original configuration and the damaged one 
are obtained via conjugate gradients minimization.
The MSD between both IS is defined as:
\begin{equation}
D^2(T,d) = \Bigl< \sum_i \left( {\bf r}_i^{IS_0} - {\bf r}_i^{IS_d} \right)^2
\Bigr >
\label{distance}
\end{equation}
in which $T$ is the temperature at which the system has been equilibrated, $d$ is the
amount of damage done to the equilibrated configuration, i.e. the number of particles
perturbed and the dislocation per particle and the braces mean an average taken over
a number of independent equilibrium configurations and by making different damages on
each configuration chosen, for example, by taking different particles to move for each
configuration.
As we want to explore the distance between $near$ IS, i.e. those accesible dynamically
at each temperature, the $damage$ should be
small, so we displace each one of the $n_d$ particles an amount $|\delta {\bf r_i}|
=0.01$ or $|\delta {\bf r_i}|=0.10$, both much smaller than the typical
interparticle distance of the system \cite{deltari}. 
Note that the distance $D^2$ goes to zero when the inherent structures of the original 
configuration and the damaged ones are the same, or in other words, when they are in the
basin of a minimum, with no double well saddles between them. 
We found that at low temperatures an important fraction
$f_0$ of the events gives zero distance,
and so in order to get sensible results we normalize $D^2$ with the fraction of non
zero events $1-f_0$. Doing this, even in the case where only one event gives a non zero 
distance, the resulting $D^2$ will be finite and equal to this distance.
Concerning the number of particles to be damaged $n_d$, 
we have checked that at low temperatures increasing $n_d$ from 2 to 40 only 
improves the statistics increasing  $1-f_0$ but not changing qualitatively 
the behaviour of $\frac{D^2(T,d)}{1-f_0}$. 
So we take $n_d$=40 and keep it fixed in the following study, the
amount of damage "$d$" in expression (1) can therefore be 
fully caracterized by the size of the dislocation $|\delta {\bf r}|$.
For temperatures $T<0.6$
we have taken averages over 1000 configurations for each one of the 32 samples
covering a time range greater than the relaxation time at each T.
Then for each configuration we have also averaged over 4 and 8 different 
particles damaged  for $d=0.10$ and $d=0.01$ respectively.

A natural question is: is there some
temperature at which $D^2(T,d)$ goes to zero? In this case this should coincide
with the point at which the order of the saddles goes to zero as discussed above. 
\begin{figure}[ht]
\includegraphics[width=7cm,height=8.5cm,angle=270]{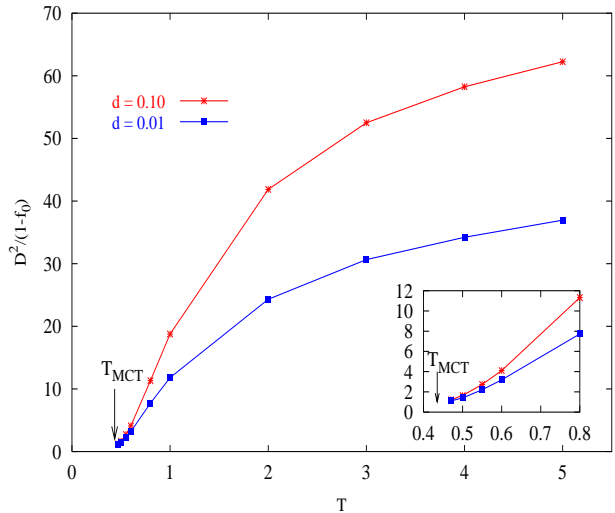}
\caption{\label{msd} Mean squared distance between inherent structures as function of
temperature for two different damages d=0.1 and 0.01. Inset: a zoom of the low
temperature region. }
\end{figure}

In figure
\ref{msd} it is shown the normalized MSD as a function of temperature for two different
damages.  As expected the distance is
very large at high temperatures and decreses upon decreasing $T$. 
Between $T=0.8$ and $T=1$ an inflection
point is clearly seen below which the MSD decreases very quickly. This temperature range
is the same at which Sastry et al.~\cite{sastry} found the onset of non-exponential
behaviour in the relaxation functions signalling the begining of what they called the
{\em landscape influenced} region. Our results are an independent confirmation of the
presence of that characteristic point but, more importantly, it gives a purely 
topological determination of it. 
From the inset it is apparent that the distance does not go to zero on
approaching $T_{MCT}$. 

\begin{figure}[ht]
\includegraphics[width=7cm,height=8.5cm,angle=270]{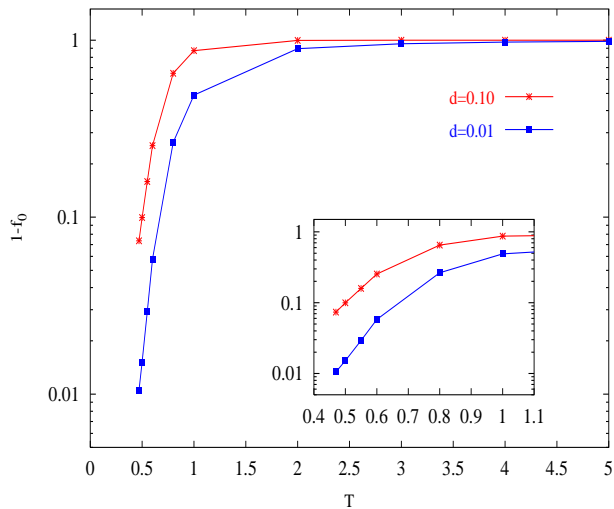}
\caption{\label{1-f0} The fraction of events with non-zero final distance as a function
of temperature for two different damages in log-linear scale. Inset: a restricted
temperature range in order to show clearly the alignment of points for $T \leq 0.6$}
\end{figure}

In order to get a better insight into 
the landscape properties of
the low temperature region we show in figure
\ref{1-f0} the probability of non zero events as a function of temperature in log-linear
scale. It is
evident the onset of a strong decrease in the non zero events at around $T=1$.
The behaviour of $1-f_0$ near $T_{MCT}$ is obviously dependent on the initial damage.
But even for the smallest damage $d=0.01$ (two orders of magnitude smaller than the
typical interatomic distance) 
the number of non zero events does not seem 
to go to zero at $T_{MCT}$.
In the inset of figure \ref{1-f0} we show $1-f_0$ in a reduced temperature range. 
The points below
$T \simeq 0.6$ are perfectly aligned indicating an exponential decay of the number of
non zero events towards $T_{MCT}$ of the form $1-f_0 =
 exp((T-T_0)/\epsilon)$ with $T_0$ depending on the initial damage. We did several
independent checks in order to confirm this behaviour. 
The direct implication of this result is that the number of saddles
does not vanish at $T_{MCT}$ in agreement with recent work~\cite{middleton,doye} 
and in apparent contradiction with previous one~\cite{angelani}. 
As we have independently verified in our simulations
that upon extrapolation to lower temperatures the diffusion constant should go to zero
at $T_{MCT}$, {\em this, together with the previous result implies 
that there exist  saddles at $T \leq T_{MCT}$ associated with nondiffusional events.}

One can get a more direct evidence of the relation between the landscape topology and
dynamics by looking directly at the particle displacements. The fact that particles
cannot diffuse in a region where there is still minima-connecting saddles available
may explain recent results obtained by B\"uchner {\em et
al.}  on the slowing down in the dynamics of supercooled liquids
on approaching $T_{MCT}$ \cite{heuer}. 
In reference\ \onlinecite{heuer}, the authors show that near $T_{MCT}$
the system visits a kind of {\em valley} containing several IS where the system is 
trapped for some time jumping between the different IS and where the mobility is
extremely small. Motion of 
particles in these  {\em valleys} should be strongly collective and probably single
displacements won't go beyond one interatomic distance. In order to test these
hypotesis we have computed several displacement distribution functions. 

\begin{figure}[ht]
\includegraphics[width=7cm,height=8.5cm,angle=270]{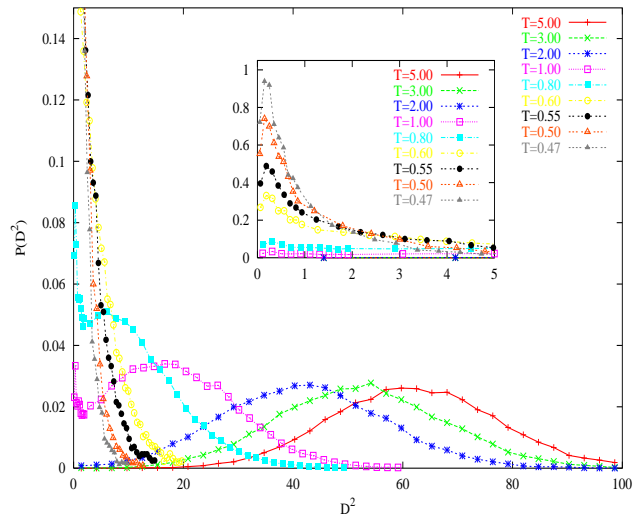}
\caption{\label{P(D2)_0:100_0:0.2}The distribution of squared displacements contributing
to the mean of equation (\ref{distance}) for the different temperatures studied
for damage $d=0.10$. Note the
presence of two maxima for T=0.8 and T=1. Inset: the peak of the distributions for
very small displacements.}
\end{figure}

\begin{figure}[ht]
\includegraphics[width=7cm,height=8.5cm,angle=270]{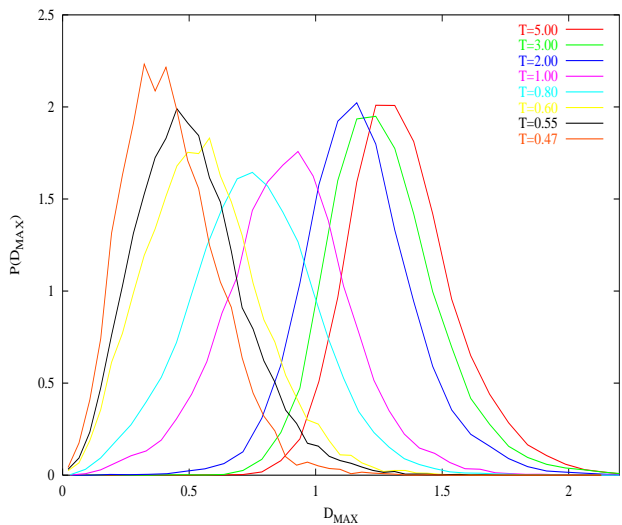}
\caption{\label{P(D_MAX)}The distribution of largest displacements for several
temperatures for damage $d=0.10$. $D_{MAX}$ is the displacement of the particle that 
experiences
the largest displacement of all the particles in the system in a given process.
Note the large dispersion in the values at $T=0.8$ and $T=1$.}
\end{figure}

In figure
\ref{P(D2)_0:100_0:0.2} we show the distribution of squared displacements, i.e. each
value of $D^2$ contributing to this histogram is a sum over the 250 particles which corresponds to
each term contributing to the mean squared displacement of equation (\ref{distance}).
We see that at high temperatures ($T > 1$) the maximum is 
located at $D^2$ greater than 40 which is the
number of particles perturbed. 
This indicates that an important fraction of the system
suffer displacements at least of the order of the interatomic distance.
An interesting fact emerges exactly at $T=1$ and is also present at $T=0.8$: a
second maximum at very low values appears. This is another indication that $T=1$ is
a characteristic temperature for this system below which the dynamics is {\em
landscape influenced} as Sastry et al. named this region. In this crossover region
three kinds of processes can be detected: 
{\em (i)} processes in which a fraction of particles can still move by
distances of the order of an interatomic distance, 
{\em (ii)} processes where the particles can only move
by tiny amounts and there is no particle moving 
an interatomic distance and 
{\em (iii)} a third group of processes where particles do
not move at all. 
The last group contributes with a delta peak at the 
origin which we have eliminated from our plots. 
The growing confinement of the particles as the temperature
is lowered from $T=1$ is more dramatic than what can be inferred from the growing
peak at low values of the distance. In fact for $T<0.8$ there is only one peak at 
$D^2 < 0.5$ (see the inset of figure \ref{P(D2)_0:100_0:0.2}). 
If we note again that each contribution to $D^2$ is a sum over the 250
particles, this implies that on average no single particle moves a distance of the 
order of
an interatomic distance. Of course, from the presence of the tails in the
distributions for larger values of the distance one cannot discard a priori the
existence of some particles that move by such amounts. Nevertheless it seems more likely
that this contributions come essentially from collective motions in which an important 
number of the particles move very little. A stronger support for this scenario is
given by the distribution of the largest displacements shown in figure \ref{P(D_MAX)}.

This figure gives also very interesting information. Note that for $T>1$ the particle that
moves more goes to a distance typically longer than one interatomic distance with very
few contributions with $D_{MAX}<1$. The variance of the distributions grow and the
height of the peak goes through a minimum as the temperature crosses the region where the landscape
begins to influence the dynamics. The large variance signals the appearance of an
important fraction of particles which, although are the ones that move more than any
other, have displacements confined to $D_{MAX}<1$. Note that the distances between IS
(figure \ref{distance}) present an inflection point between $T=0.8$ and $T=1$ indicating 
a rapid decrease in the number of saddles available for diffusing. 
For $T<0.8$ the maximum displacements
are peaked around very small values less than one interatomic distance.

\begin{figure}[ht]
\includegraphics[width=7cm,height=8.5cm,angle=270]{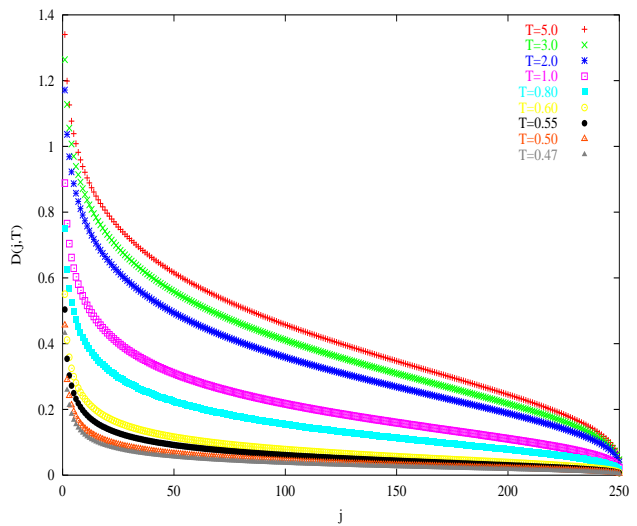}
\caption{\label{pattern}The pattern of displacements for the 250 particles and for
several temperatures}
\end{figure}

Finally we asked the question: is there some order in the pattern of displacements per
particle as a function of temperature? A partial but important answer can be done by
ordering the N particles by decreasing value of their displacements. In figure
\ref{pattern} we show that in fact the patterns of displacements are 
similar for all temperatures. 

The first point to the left corresponds, for each temperature,
 to the mean over the particles with the largest displacements
 (first moment of distributions of figure \ref{P(D_MAX)}), 
 the second point is
the mean over particles with the second largest displacements, and so on up to the
particles with the smallest displacement. From this figure it is very clear that the
overwhelming majority of particles move in average less than 1 while only very few move more than
one interatomic distance at high temperatures $T>1$. For $T<0.8$ all particles in average
move very little. Interestingly, the average distance traveled by the particles with the
largest displacement goes to zero very slowly (almost linearly)
as $T\rightarrow 0$ suggesting again a continuous variation
of landscape properties on crossing $T_{MCT}$.
Also note that most particles move by some amount which means that
the motion is highly collective. By looking at the relative difference between the
first and second largest displacements (not shown) we obtained a practically constant
and low value for $T > 1$ followed by a pronounced grow of
this difference for $T < 1$ which is another indication of the drastic confinement of
the system in this range of temperatures where only one particle moves much more than
the others (although in a scale less than one interatomic distance).

Recently Grigera et al.
\cite{grigera} computed the distribution of particle displacements between a minimum
and a saddle-connected neighbour in a soft spheres system and found evidence that,
at low temperatures, the typical process is one
where a large amount of particles
travel very small distances, less than half an interatomic distance, and
very few particles move by the order of one interatomic distance.
These were interpreted as activated processes in a region
where there were almost no saddles available.
This picture is qualitatively different from the one that
emerges
from the present work for 
the binary LJ system. 
Nevertheless,
further work is needed to elucidate if diffusion in
binary LJ system occurs as a cumulative effect 
of the small collective displacements or
it is due to very rare activated events that,
if present, should be confined to the tails of our
displacement distributions.

In summary we analyzed the information brought by the distances between near inherent
structures relevant for the dynamics of a supercooled liquid near the mode coupling
transition temperature. In this way the contributions coming from vibrations are
automatically filtered out. It is possible to obtain rather precise information on the
processes involved in the evolution in phase space and also in real space. The functional
form of the MSD as a function of temperature and the evolution of the distribution of
displacements and of maximum displacements define clearly a characteristic region near
$T=1$ below which the number of saddles decays rapidly and the particles become strongly
confined. We found also that the number of saddles is exponentially small on approaching
the mode coupling transition temperature but does not go to zero.
In the low temperature region, in connecting two neighbouring IS,  most of the times all 
particles move by amounts much
smaller than the typical interparticle distance. The scenario that emerges for the low 
temperature dynamics is one in which the landscape is formed by a kind of metabasins 
\cite{metabasin,middleton,doye} 
in which there are
still many inherent structures connected by low lying saddles which may exist even for
$T<T_{MCT}$ (see also recent work by Heuer {\em et al.} \cite{heuermeta}). Individual particles move by tiny amounts within a metabasin in a highly
collective way. 
From our results on the binary LJ system the mode coupling transition temperature $T_{MCT}$
continues to be characterized by a
dynamical singularity and the possible connection of the transition itself with a sharp 
change in some topological property of the potential energy landscape 
remains to be elucidated.

\begin{acknowledgments}
We thank Tomás Grigera and Andrea Cavagna for interesting comments.
This work was partly supported by CONICET and {\em Fundaci\'on Antorchas}, 
Argentina and by CNPq and FAPERGS, Brazil. 
\end{acknowledgments}

\end{document}